\documentclass[12pt]{elsarticle}
\usepackage{graphicx}
\usepackage[utf8]{inputenc}  
\usepackage[T1]{fontenc}
\usepackage{amsfonts,amssymb,amsthm,amsmath}
\usepackage{subfigure}
\usepackage[french,english]{babel}
\usepackage{graphics,graphicx}
\begin{document}

\title{Analysis of the Hubble diagram of type SNe Ia supernovae and of $\gamma$-ray bursts. A comparison between two models}

			\author{J.-M. Vigoureux$^{*}$, D. Vigoureux$^{**}$, P. Vigoureux$^{*}$, M. Langlois$^{*}$}
					 
			\address
				{$^{*}$
				 Institut UTINAM, UMR CNRS 6213,\\
				 Universit\'e de Franche-Comt\'e, 25030 Besan\c{c}on Cedex, FRANCE\\
				  {\em jm.vigoureux@free.fr}\\
				  {\em jean-marie.vigoureux@univ-fcomte.fr}}
			\address
				{$^{**}$
				 Lyc\'ee Jules Ferry,\\
				 29 rue du Mar\'echal Joffre, 78000 Versailles, FRANCE\\
				 {\em dorian.vigoureux@ac-versailles.fr}}

\begin{abstract}
\noindent A paper by Harmut Traunmüller \cite{HT} showed from statistical studies of observational data that the most adequate equation to represent observations on magnitude and redshift from 892 type 1a supernovae is  $\mu = 5\,log[(1+z)\,ln(1+z)] + const.$\\  
Comparing the Hubble diagram calculated from the observed redshift data of 280 supernovae with Hubble diagrams inferred on the basis of two cosmological models in the range of z = 0.0104 to 8.1, Laszlo Marosi \cite{Marosi} found in a quite independant study that the best fit function to represent observations is 
$\mu= 44.109769 \,z^{0.059883}$. Noting that differences between the different cosmological models become more pronouced in a photon time-of-fligth $t_s$ \textit{vs}. $z$ représentation, he also noted that the best equation to account for observations may also be written $z = -1+e^{2.024\,\,10^{-18} \,\,ts}$.\\
In the light of these \textit{observational} data, we compare 
the theoretical Hubble diagram obtained with the flat $\Lambda CDM$ model to the ones we have obtained few years ago \cite{Vigoureux08, Viennot09, Vigoureux} from a model that we call here the "light model" for the sake of clarity. \\ Our conclusions are that
values calculated on the basis of the $\Lambda CDM$ model exhibit poor agreement with the presently available data while the light model agrees \textit{exactly} with observations and conclusions of statistical studies \cite{HT} and \cite{Marosi} (independently of the values of $\Omega_k$, $\Omega_M$ or $\Omega_{\Lambda}$). Our model giving no accelerating expansion of the universe, we conclude that this latter is not necessary and that models can exist which lead exactly to observations without having to consider any accelerating expansion of the universe. In an Appendix, we discuss some aspects of the model and we present a brief overview of some of its key results.\\

\end{abstract}
\maketitle \fontsize{12}{24}\selectfont

\section{Introduction}
Following pioneering works related in  \cite{Norgaard-Nielsen89}, observations of type Ia supernovae \cite{Riess98, Perlmutter99,Tonry03,Wang03,Riess04,Schwarzschild04} have provided a robust extension of the Hubble diagram. These results have shown that
observations cannot be fitted by using the usual distance modulus expression with $\Lambda = 0$ both for $z < 1$ and for $z > 1$. To fit new data points at redshift 1.755 the standard model thus needs to consider that the expansion of the universe is accelerating, an effect that is attributed to the existence of a dark energy.\\
The luminosity distance \textit{vs}. redshift law has also been measured using $\gamma$-ray bursts which have given a complementary probe to SN Ia (see \cite{Sch} and ref. therein). 
Taking into account all measurements,  the Hubble diagram is now continuously sampled in the range of $z= 0.1$ to $8.1$.\\

\begin{figure}[h]
\centering
\subfigure[]{\includegraphics[scale=0.4]{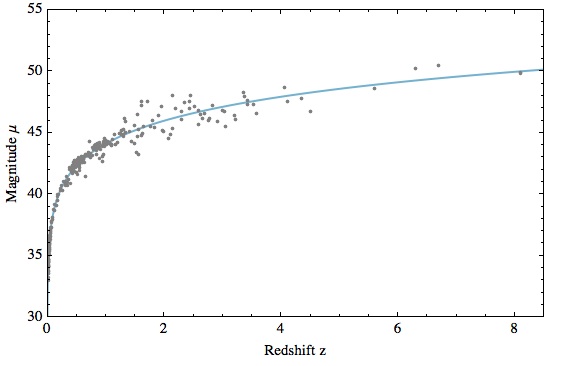}}
\subfigure[]{\includegraphics[scale=0.4]{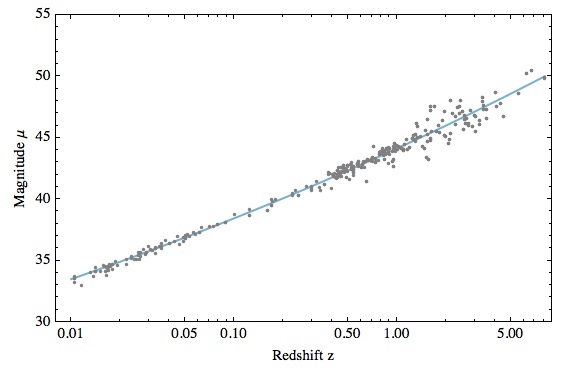}}
\caption{
The distance modulus $\mu$ \textit{vs.} redshift $z$ from 557 SNIa and 50 $\gamma$-ray bursts. Data points are taken from Riess and coauthors \cite{Riess04} and from Wei \cite{Wei10} but we don't represent here all data points taking only some of them for illustration. a) The full line curve represents the best fit of observations as found by H. Traunmüller  (eq.\ref{lum}) \cite{HT} and L. Marosi (eq.\ref{muMarosi}) \cite{Marosi}. As we shall see in what follows, it also corresponds exactly to results obtained when using the "light model". b) The same as fig.(1a) but now plotting $\mu$  \textit{vs.} log(z)  \label{fig 1}}.
\end{figure}

Analysing these results,
Harmut Traunmüller \cite{HT}, using observational data on magnitude and redshift from 892 type Ia supernovae, showed that for standard candles, "\textit{magnitude $\mu = 5\,log[(1+z)\,ln(1+z)] + const.$ gives the best fit of all results}" so that
\begin{equation} \label{HT}
\mu = \text{Const.}  +5 \log{ \Big((z+1) \ln{(z+1)}\Big)} 
\end{equation}
Statistical regression analysis permits to calculate the constant appearing in that equation so that the best fit of observations on the whole range of $z = 0.01$ to $8.1$ may be written
\begin{equation}\label{lum}
\mu = 43.3856557+ 5 \log{ \Big((z+1) \ln{(z+1)}\Big)} 
 \end{equation}
For his part, Laszlo Marosi \cite{Marosi, Marosi14, Marosi16} found quite independently and also from statistical considerations, that the best fit function obtained from data of 276 supernovae in the range of $z= 0.01$ to $8.1$ is:
\begin{equation}\label{muMarosi}
\mu= 44.109769 \,\,z^{0.059883}    
\end{equation}
These two \textit{independent} results (\ref{lum}) and (\ref{muMarosi}) give the same numerical values at a relative difference $\frac{eq.(\ref{lum}) - eq.(\ref{muMarosi})} {eq.(\ref{muMarosi})}
 \leqslant \,0.01 \% $. 
We thus consider in what follows that they are the same and we call eqs.(\ref{lum}) and (\ref{muMarosi}) "experimental results" or "observations". \\
Noting that differences between different cosmological models become more prononced in a photon flight time $t_s= \frac{D_c}{c}$ versus $z$ representation ($D_c$ is the co-moving radial distance), Laszlo Marosi has also showed \cite{Marosi14} that the best statistical fit of observational data may be written 
\begin{equation}\label{M}
z = -1+e^{2.024\,\,10^{-18} \,\,ts}
\end{equation}
In the first part of this paper, our aim is to compare theoretical results of the standard flat $\Lambda CDM$ model with results (\ref{lum}-\ref{muMarosi}) and (\ref{M}) which may be considered as the best representations of observations.
We thus show that results calculated on the basis of a flat $\Lambda CDM$ model exhibit poor agreement with the observational data.\\
In a next part, we come back to the "light model" we have presented few years ago \cite{Vigoureux08, Viennot09, Vigoureux, Vigoureux14} and we show that this latter leads exactly to eqs(\ref{lum}, \ref{muMarosi}, \ref{M}) and thus provides a \textit{perfect} match with observation. This model leading to experimental observations without having to consider any expansion of the universe, we deduce that accelerating expansion (which is necessary in the standard model), is not necessary in that model.
We come back in an Appendix on the main features of the light model.

\section{Comparison of \textit{theoretical} results obtained from the standard flat $\Lambda CDM$ model with \textit{observations}}
\subsection{Comparison of standard results with observations (eqs.\ref{lum}-\ref{muMarosi})}
Our aim is to compare the Hubble diagramm inferred on the basis of a flat $\Lambda CDM$ model to the one obtained from observations and expressed by eqs.(\ref{lum}-\ref{muMarosi}).
In the standard model, the distance modulus can be computed from:
\begin{equation}\label{mucl}
\mu = 25 + 5 \log{\Big((1 + z) \int\limits_{0}^{z} \frac{c\, dz'}{H_0 \sqrt{(z'+1)^3
  \, \Omega_M+(z'+1)^2\, \Omega_k+\Omega_{\Lambda}}}\Big)}
\end{equation}
where $H_0 = H(t_0)$ is the Hubble constant at the present time $t_0$.\\
As explained above, although obtained independently through the use of statistical regression techniques, eqs.(\ref{lum}-\ref{muMarosi}) lead to the same numerical values at less than $0.01\, \%$. The comparison between the theoretical result (\ref{mucl}) and observations may then be made by using any of these two equations (\ref{lum}-\ref{muMarosi}). \\
Figs. 2a and 2b are obtained in the case of the current picture of a flat universe $(\Omega_k \simeq 0$). They show that the flat $\Lambda CDM$ model cannot exactly account for observations in the whole range $0.1\,<\,z\,<\,8.1$: when $\Omega_M$ and $\Omega_{\Lambda}$ are chosen in order to account for observations in the range of small $z$ values $z\,< 2$ ($\Omega_M = 0.27$ and  $\Omega_{\Lambda}= 0.73$) they cannot explain observational data for higher values of $z$ (and conversely). \\
\begin{figure}[h]
\centering
\subfigure[]{\includegraphics[scale=0.4]{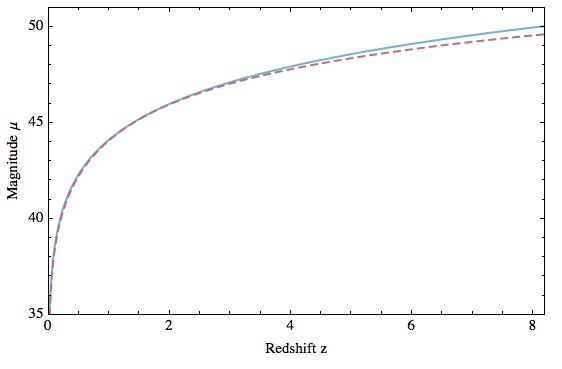}}
\subfigure[]{\includegraphics[scale=0.4]{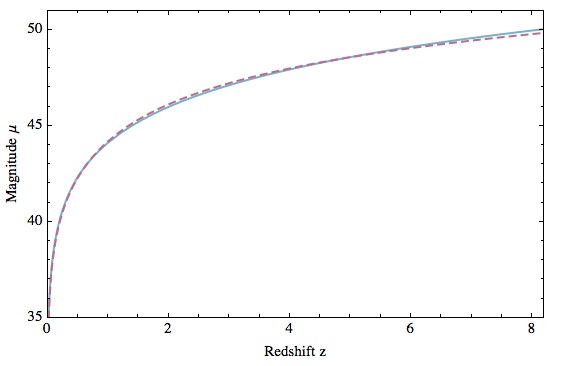}}
\caption{Figs (2a) and (2b) compare the variations of the distance modulus $\mu$ vs. $z$ in the range $0\,< \,z \,< \,8.1$ obtained from the flat $\Lambda CDM$ model (dashed line) to experiments (full line).
They are both obtained with $H_0\simeq 73\, \text{km}\,\text{s}^{-1} \text{Mpc}^{-1}$ (but a small change of that value does not change significantly main results). Fig.2a shows that when $\Omega_M$ and $\Omega_{\Lambda}$ are chosen in order to account for observations in the range $0.15\,< \,z\,< 3$ (fig.2a is obtained with $\Omega_M = 0.27$ and $\Omega_{\Lambda}= 0.73$), they cannot account for observations for higher values of $z$. Fig.2b shows that when $\Omega_M$ and $\Omega_{\Lambda}$ are adapted to get a better agreement for $z$ values around $4-5$ (fig.2b is obtained with $\Omega_M = 0.20$, and $\Omega_{\Lambda}= 0.80$), they cannot account for observations for smaller and higher values of $z$. Note that eqs.(\ref{lum}-\ref{muMarosi}) (full lines) giving the best statistical fit of observations, it is useless to represent here data points as in fig (1).}.
\end{figure}
We could think that this discrepancy could come from the fact that all the data in the range $0.1\,<\,z\,<\,8.1$ have not been obtained in the same conditions but that is not the reason. It can in fact be shown (fig. 3) that this same discrepancy remains in a more restricted range such as $0.1\,<\,z\,<\,0.9$.\\
\begin{figure}[!h]
\centering
\includegraphics[scale=0.45]{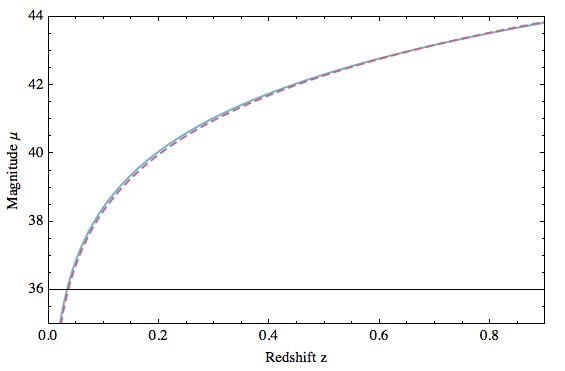}
\caption{The variations of the distance modulus $\mu$ vs. $z$ in the narrower range $0\,< \,z \,< \,0.9$.
The best fit of observations (eqs.(\ref{lum}-\ref{muMarosi})) 
is in full line when the curve corresponding to the flat $\Lambda CDM$ model with $\Omega_M = 0.27$ and $\Omega_{\Lambda}= 0.73$ is in dashed lines. It shows that the curve obtained with eq.(\ref{mucl}) is perceptibly different from observations. We conclude that, in the smaller range $0.1\,<\,z\,<\,0.9$ of $z$ values as in its whole range $0.1\,<\,z\,<\,8.1$ (figs. 2a and 2b) a given set of $(\Omega_M,\,\Omega_{\Lambda})$ values cannot account for \textit{all} observations.\label{fig 3}}
\end{figure}
So, a given set of values of $\Omega_M$ and $\Omega_{\Lambda}$ cannot express exactly observations neither over the entire range of $z= 0.01$ to $z = 8.1$ nor in a narrower range as $z= 0.01$ to $z = 0.9$. The set of values $(\Omega_M,\,\Omega_{\Lambda})=(0.27,\,0.73$), which leads to good experimental values for small $z$ is no longer suitable for higher values of $z$, and, conversely, those (for example $(\Omega_M,\,\Omega_{\Lambda})=(0.20,\,0.80$)) which would account for observations for the higher values of $z$ correspond no more for small values of $z$.\\ 
The same conclusion can also be found by calculating the difference between the distances modulus obtained from observations $\mu_{obs}$ (eq.\ref{lum}) and the ones obtained from the $\mu_{\Lambda CDM}$ model (eq.\ref{mucl}). This difference is plotted on fig.4 which shows that when inserted in eq.(\ref{mucl}), the values $\Omega_M = 0.27$, $\Omega_{\Lambda}=0.73$ lead to the good experimental values for $z$ around $1.6$ but not elsewhere. \\

\begin{figure}[!!h]
\centering
\includegraphics[scale=0.5]{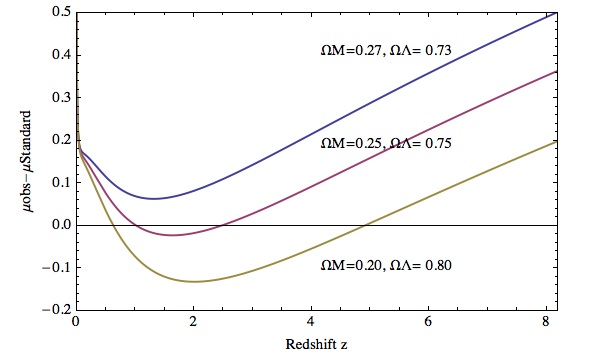}
\caption{variations  with respect to $z$ of the difference $\Delta\mu= \mu_{obs} - \mu_{\Lambda CDM}$  between observations and the theoretical results obtained from the flat $\Lambda CDM$ model. The horizontal line $\Delta\mu= \mu_{obs} - \mu_{\Lambda CDM}\,=\,0$ corresponds to an exact match between results of the flat $\Lambda CDM$ model and observations.
The values $(\Omega_M,\,\Omega_{\Lambda})=(0.27,\,0.73$) give a good result for $z$ around $1-2$ but not elsewhere (the difference $\Delta\mu$ is in fact around $0$ near $z = 1-2$ but it increases for greater or lower values). The same is true for other values of $(\Omega_M,\,\Omega_{\Lambda})$: they cannot account for all observations in the whole range of $z$ values.
\label{fig 4}}
\end{figure}

\indent Of course, there are a large number of different cuts through the 4 dimensional parameter space ($z,\,H_0,\, \Omega_M,\,\Omega_{\Lambda}$) that can be plotted and we can ask ourselves if it could be possible to find other values of ($\Omega_M$, $\Omega_{\Lambda}$) and of $H_0$ to get a best agreement between observations and the standard theory.\\
\indent To answer that question we present in fig.5a the contour in the ($\Omega_M-z$) plane which corresponds to an exact concordance between the flat $\Lambda CDM$ model and observations. The curves still clearly show that a given value of $\Omega_M$ (and therefore of $\Omega_{\Lambda} = 1\,-\,\Omega_M$) cannot lead to observed values for all $z$. \\
The same question can be asked to find a best value of the Hubble constant $H_0$ which would reproduce
observations for a given set of $\Omega_M$ and $\Omega_{\Lambda}$ values. This leads us to the same conclusion (fig.5b): the value of $H_0$ which exactly corresponds to observations for a given value of $z$ does not correspond anymore for other values. For example, the value $H_0 \sim 70 $ corresponds to observation for $z \sim 1.7$ but redshift around $z\sim 6$ can only be found with $H_0 \sim 62$\\

\begin{figure}[h]
\centering
\subfigure[]{\includegraphics[scale=0.4]{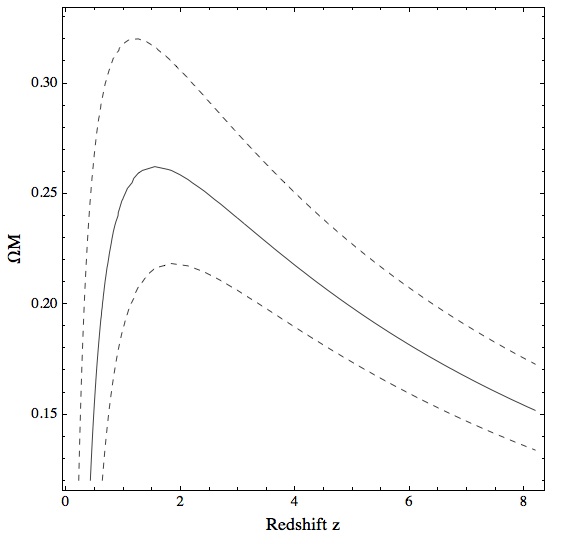}}
\subfigure[]{\includegraphics[scale=0.4]{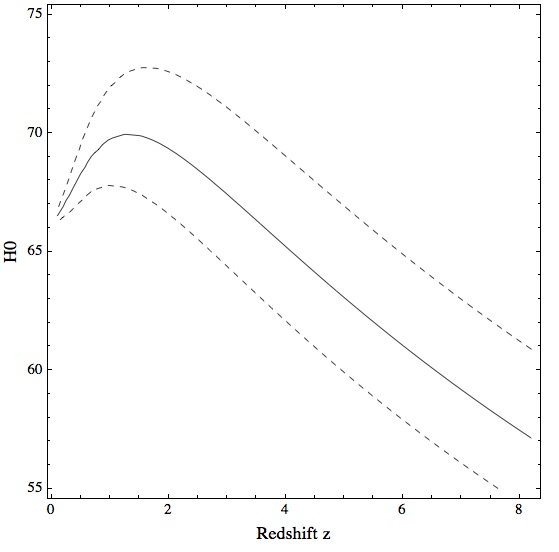}}
\caption{Fig.5a is drawn in the ($\Omega_M-z$) plane. It shows the values of $\Omega_M$ which \textit{exactly} give the best fit of observations for each $z$ in a flat $\Lambda CDM$ model with $H_0= 73\,\text{km}\,\text{s}^{-1} \text{Mpc}^{-1}$. We see that the same value of $\Omega_M$ cannot account for observations whatever may be $z$. The dashed lines show results obtained with $H_0= 69\,\text{km}\,\text{s}^{-1} \text{Mpc}^{-1}$ (the upper curve) and $H_0 = 75 \,\text{km}\,\text{s}^{-1} \text{Mpc}^{-1}$ (the lower curve). They show that changing $H_0$ cannot give better results. Fig.5b is drawn in the ($H_0-z$) plane. It shows the same result in terms of $H_0$ and $z$ in a flat $\Lambda CDM$ universe with $\Omega_M=0.27$ and $\Omega_{\Lambda}= 0.73$. The dashed curves are obtained with $(\Omega_M,\, \Omega_{\Lambda})\,=\,(0.25,\,0.75) $ (upper curve) and $(\Omega_M,\, \Omega_{\Lambda})\,=\,(0.35,\,0.65)$ (lower curve).}
\end{figure}

\section{The light model}
The above analysis shows that results calculated on the basis of a flat $\Lambda CDM$ model exhibit poor agreement with observational data.\\
Our aim now is to  come back to another cosmological model (that we call here the "light model" for the sake of clarity) that we have proposed some years ago \cite{Vigoureux08} \cite{Viennot09} \cite{Vigoureux}, to show that this latter expresses \textit{exactly} observations (\ref{lum}), (\ref{muMarosi}) and (\ref{M}) \textit{over the entire range of $z$ values}.\\
Noting that both $c$ and
the expansion of the universe provide a universal relation between space and time which both have
the physical dimension of a velocity, we consider that these two facts cannot be a fortuitous
coincidence and that they are two different aspects of a same phenomenon.
We have thus proposed \cite{Vigoureux08, Vigoureux88} that the constant "$c $" (the so-called speed of light) and the expansion of the universe are two aspects of one single concept connecting space and time in the expanding universe by putting (we write the scale factor in a normalized form $a(t)= R(t) /R_0$ and the subscript $0$ always refers to a quantity evaluated at the present time $t_0$): 
\begin{equation} \label{c}
c = \alpha \frac{dR(t)}{dt} =\alpha \,R_0\, \frac{da(t)}{dt} = Const.
\end{equation}
where, $c$ and $\alpha$ are constants (let us underline that, if a zero mass density, zero pressure and zero cosmological constant imply $\ddot{R}(t)=0$, the reverse is not true: $\ddot{R}(t)=0$, as obtained from eq.\ref{c}, does not imply that these physical quantities are separately zero if $\Lambda$ is allowed to vary as a function of time (see Appendix, eq.\ref{Friedc2}).
 Moreover we will show in the appendix that eq.(\ref{c}) is compatible with Friedmann equations and with the conservation law of energy).\\
It can also be noted that eq.(\ref{c}) permits to define $c$ from the knowledge of the geometry of space-time only, that is from its size and its age and thus really gives c the statute of a true geometrical fundamental magnitude of the universe.\\
Our first aim is to show that eq.(\ref{c}) leads exactly to observations as modelled by eqs.(\ref{lum}, \ref{muMarosi}, \ref{M})

\subsection{Comparison of \textit{theoretical} results obtained from the light model with observations}
In that part, we show that eq.(\ref{c}) leads to an expression for the distance-moduli $\mu$ which can fit \textit{exactly} all the data.\\
To calculate the expression for the distance modulus with respect to $z$, let us consider an 
object at cosmic coordinate $\chi$ and let us suppose that the light that is emitted at cosmic time $t_e$ is just reaching us at time $t_0$. Let us also write the Robertson-Walker metric in the form
\begin{equation}
ds^2=- c^2 dt^2 + R(t)^2 \left(d\chi^2 + S_k^2 \,d\Omega^2 \right) 
\end{equation}
where $R(t)=R(t_0)\, a(t)$. Using (\ref{c}), which obviously gives $R(t_0) \equiv R_0= c \,\dfrac{t_0}{\alpha}$, $H_0= \dfrac{\dot{a}(t_0)}{a(t_0)} = \dfrac{1}{t_0}$ so that $R_0=\dfrac{c}{H_0\, \alpha}$, the luminosity distance $d_L$ of the object can be written 
\begin{equation}\label{dL1}
d_L=(z+1)\,R_0 \, \chi= \frac{c}{H_0 \, \alpha}\,(z+1)\, \chi
 \end{equation}
$\chi$ can be obtained by writing that light travels on a radial null geodesic so that $c\, dt = R(t)\, d\chi$. Using (\ref{c}) and integrating $d\chi$ between $\chi_e$ ("e" for emission time) and $\chi_0 = 0$ ($\chi =0$ at the present time) we get:
 \begin{equation}
\chi_e =\int_{t_e}^{t_0}\frac{c\, dt}{R(t)} =\int_{t_e}^{t_0}\frac{\alpha\, da(t)}{a(t)}= \alpha \ln{\frac{a(t_0)}{a(t_e)}}
\end{equation}
Introducing this result into eq.(\ref{dL1}) with $a(t_0)= a(t_e) (z + 1) = 1$, gives
\begin{equation}\label{dL}
d_L=\frac{c}{H_0}(z+1) \ln{(1 + z)}
\end{equation}
The distance modulus being related to the luminosity distance via
\begin{equation}
\mu = 5\, \log{\left(d_L(Mpc)\right)}+25
\end{equation}
using (\ref{dL}) this result gives
\begin{equation}\label{muV}
\mu = 25 + 5\, \log{ \left(\frac{c}{H_0}\right)}+5 \log{ \Big((z+1) \ln{(z+1)}\Big)} 
 \end{equation}
 where $c$ is in \text{km.s}$^{-1}$ and $H_0$ in $\text{km}\,\text{s}^{-1}\, \text{Mpc}^{-1}$.\\
 Eq.(\ref{muV}) which we found in \cite{Vigoureux08} is exactly the one (eq.(\ref{lum})) that Harmut Traunmüller \cite{HT} found quite independently from statistical analysis as being the best to interpret the observations on the whole range $0.1\,<\,z\,<\,8.1$ of $z$ values. Having previously shown that results of Traunmüller (eq.(\ref{lum})) and those of Marosi (eq.(\ref{muMarosi})) are quantitatively  the same,
we may conclude that using eq.(\ref{c}) succeeds in explaining all the experimental data in the whole range of $z$ values.
It is therefore useless to draw a figure to compare the result (\ref{muV}) with observations: the full line in figs.(1, 2 and 3) which represents the best fit of observations as found in \cite{HT} and \cite{Marosi} also corresponds to the theoretical result (\ref{muV}) obtained from eq.(\ref{c}).\\
	Comparing eq.(\ref{muV}) and eq.(\ref{lum}) permits to calculate the Hubble constant $H_0$. Writing 
$$   25 + 5\, \log{ \left(\frac{c}{H_0}\right)}\,=\,43.3856557 $$
leads to $H_0=63\, \text{km}\,\text{s}^{-1} \,\text{Mpc}^{-1}$. This value may seem rather low when compared to the standard value $73.02 \pm 1.79\;\text{km}\,\text{s}^{-1}\, \text{Mpc}^{-1}$ \cite{Riess16} obtained from SNe Ia supernovae and Cepheids or even to the value $67.27 \pm 0.66\,\text{km}\,\text{s}^{-1}\, \text{Mpc}^{-1}$ predicted by Planck Collaboration and \textit{al.} \cite{Planck2015}, but the following points must be noted:\\
The statistical fits (\ref{lum}, \ref{muMarosi}) of observations on the whole range of $z$ values have been obtained from \textit{independant} studies which both imply a value around $H_0 \sim 63 \,\, \text{km}\,\text{s}^{-1}\, \text{Mpc}^{-1}$ (moreover we will see in what follows how eq.(\ref{M}) also leads to that same value of $H_0$). The value $H_0\simeq 63\,\, \text{km}\,\text{s}^{-1}\, \text{Mpc}^{-1}$ so appears to be the value which may account for observations on the whole range $0.1\,<\,z\,<\,8.1$ \textit{in the context of the light model}. 
These latter words are to be underlined. It must in fact be noted that the value of $H_0$ is never measured \textit{directly}. It can only be calculated \textit{in the field of a given theory}. As liked to recall Einstein, theory and observations are interdependent and there is no observation which can be directly interpreted by itself without refering to a given theory. It thus depends in a fundamental way on the cosmological model we consider. When inserted in the $\Lambda CDM$ model, observations lead to $H_0 \sim 73$. This does not imply that any other theory must find that value. \\
To be clear: the same observations can lead to different interpretations and different numerical results when interpreted within different theories. A simple example of this can be given by the deceleration parameter $q$: \textit{the same observations} lead to a value close to $-0.5$ for today and close to $+ 0.5$ for very high redshifts when they are interpretated in the context of the prevailing $\Lambda CDM$ model, whereas they lead to $q=0$ at all times when interpretated with the light model ($c$ being constant in eq.(\ref{c}) the light model implies  $\ddot{R}(t)=0$).\\
	\indent It is also interesting to see that (\ref{muV}) is independant of $\Omega_M$ and $\Omega_{\Lambda}$ so that the distance modulus here appears as a pure metric quantity.

\subsection{Comparison of results obtained from the light model with eq(\ref{M})}
\noindent It is easy to show that eq.(\ref{c}) also leads to observations (eq.\ref{M}) as fitted in \cite{Marosi14}. \\
Using successively $c = \alpha\,\dot{R}(t)= \alpha\,\dot{R}(t_0)=Const.$ and $\dot{R}(t_0)=H_0 \,R(t_0)$, the photon flight-time $t_s$ can be written:\\
\begin{equation}
t_s=\frac{R_0\,\chi}{c} =\frac{R_0 \chi}{\alpha \dot{R}_0}=\frac{\chi}{\alpha \,H_0}
\end{equation}
integrating then 
\begin{equation}
d\chi=-\frac{c\, dt}{R(t)} = -\frac{\alpha \,da(t)}{a(t)}
\end{equation}
between $\chi_e = \chi$ and $\chi_0=0$ along a light ray gives (we also use $a(t_0) = (1+z) \,a(t)$)
\begin{equation}\label{tslum}
t_s=\frac{\text{ln}(1+z)}{H_0}
\end{equation}
and consequently
$$z\,=\,-1+e^{t_s\, H_0}$$
Comparing this result with eq.(\ref{M}) gives $H_0= 62.5 \,\text{km}\,\text{s}^{-1}\, \text{Mpc}^{-1}$ which is very near the value $H_0= 63 \,\text{km}\,\text{s}^{-1}\, \text{Mpc}^{-1}$ found above from another analysis.\\
To conclude, all the data are consistant with the "light model" which can express exactly the three independant fits of observations (\ref{lum}-\ref{muMarosi}) and (\ref{M}) over the entire range of $z= 0.01$ to $z = 8.1$

\subsection{The origin of the difference between results obtained with the light model and those obtained with the standard $\Lambda CDM$ theory}
We may ask why eq.(\ref{c}) directly leads to the "best fit of experiments" as obtained by Marosi and Traunmüller.\\
To answer this question, our aim is to explain the difference between eq.(\ref{muV}) and the usual one (\ref{mucl}) and to show why eq.(\ref{muV}) does not depend on $\Omega_M$, $\Omega_{\Lambda}$ and $\Omega_k$. We will see that this difference comes from the fact that when using eq.(\ref{c}) 
the cosmological "constant" $\Lambda$ and the matter density $\rho(t)$ both vary with time as $a(t)^{-2}$ whereas in the standard model $\Lambda$ is constant and $\rho(t)$ varies with time as $a(t)^{-3}$. \\
The general expression for the luminosity distance $d_L$ can be expressed in terms of an integral over the redshift $z'$ of the propagating photon as it travels from $z'$ to us at $z'=0$. We have:
\begin{equation}\label{dldl}
d_L = (1+z)\, R_0\, \chi = (1+z)\int_0^z \frac{c\, dz'}{H_0\, E(z')}
\end{equation}
where
\begin{equation} \label{Pee}
E(z) =\sqrt{\Omega_M (1+z)^3 + \Omega_k (1 + z)^2 + \Omega_{\Lambda}}
\end{equation}
In order to calculate $E(z)$ in the more general way,  let us consider that the matter density $\rho_M(t)$ and the density of cosmological constant $\rho_{\Lambda}(t)$ vary as $a(t)^{-p}$ and $a(t)^{-q}$ respectively so that
\begin{eqnarray}\label{ro}
\rho_M (t) &=& \rho_M (t_0)\,\frac{a(t_0)^p}{ a(t)^{p} }= \Omega_{M} \,\rho_c\,\frac{a(t_0)^p}{ a(t)^{p} }  \nonumber\\
\rho_{\Lambda}(t)&=& \rho_{\Lambda} (t_0)\, \frac{a(t_0)^q}{a(t)^q}= \Omega_{\Lambda}\,\rho_c\, \frac{a(t_0)^q}{a(t)^q}
\end{eqnarray}
where $\rho_c$ is the critical density.
Introducing (\ref{ro}) into the first Friedmann equation (see Appendix) gives:
\begin{eqnarray}\label{Friedmann}
H(t)^2&=&\frac{8 \pi  G}{3}  (\rho_M (t)+\rho_{\Lambda} (t))-\frac{c^2 k}{a(t)^2 R_0^2}
\end{eqnarray}
Writing sucessively this result at times $t$ and $t_0$ with $\Omega_{k0} = - \dfrac{c^2 k}{H_0^2\,a(t_0)^2)\,R_0^2}$ and $\rho_c=\dfrac{3 H_0^2}{8 \pi G}$ we obtain the two equations:
\begin{equation}
H(t)^2=H_0^2 \left(\Omega_{M} \frac{a(t_0)^p}{a(t)^{p}} +\,
   \Omega_{k} \frac{a(t_0)^2}{a(t)^2}+\,
   \Omega_{\Lambda} \frac{a(t_0)^q}{a(t)^{q}} \right)\nonumber
  \end{equation}
  \begin{equation}\label{1}
1 =\Omega_{M}+ \, \Omega_{k}+\,\Omega_{\Lambda}
\end{equation}
so that using
$$ a(t)=\frac{a(t_0)}{1+z}$$
we get the general result $H(t)=H_0 \,E(z)$ with
\begin{equation}\label{Ez}
E(z) =\sqrt{\Omega_{M} (z+1)^p\,+\,\Omega_{k}\,(z+1)^2 +\Omega_{\Lambda}(z+1)^q\,}
\end{equation}
In the standard case ($p = 3, \,q = 0$ or $\rho_M\, \sim \,a^{-3}, \rho_{\Lambda}\, \sim \,a^0$), this result gives\\
 \begin{equation}\label{Ezst}
E(z) =\sqrt{\Omega_{M} (z+1)^3\,+\,\Omega_{k}\,(z+1)^2+\,\Omega_{\Lambda}}
\end{equation}
which is used in eq.(\ref{mucl}), whereas in the light model  $(p = 2,\, q=2$ or $\rho_M \, \sim\, a^{-2}, \rho_{\Lambda}\,\sim \,a^{-2}$, (see the Appendix) eq(\ref{Ez}) gives
 \begin{equation}
E[z] =\,(1+z)\sqrt{\Omega_{M}\,+\,\Omega_{\Lambda}\,+\,\Omega_{k}}
\end{equation}
so that using the secund eq.(\ref{1}) gives
 \begin{equation}\label{Ezlum}
E(z) = (1+z)
\end{equation}
Using this result into (\ref{dldl}) gives (\ref{dL}).
We thus understand the fundamental origin of the difference between the standard model and the light model.\\

This above result also explains the difference between eq.(\ref{tslum}) which corresponds to experimental results as expressed by Marosi (eq.(\ref{M})) and the one obtained from the flat $\Lambda CDM$ model.
In a general way, the photon flight-time can be obtained from
\begin{equation}\label{ts}
 t_{s}(z) = \int_{0}^{z} \frac{dz'}{H_0\, E(z')}
\end{equation}
In the $\Lambda CDM$ model $E(z')$ is given by eq.(\ref{Ezst}) whereas it is given by (\ref{Ezlum}) in the light model. Introducing this latter result into (\ref{ts}) then leads to (\ref{tslum}) (and consequently to observations).\\
We may be surprised by a variation of $\rho_M(t)$ and $\rho_{\Lambda}(t)$ as $a(t)^{-2}$ in our equations. Such a result remains of course to be discussed. However, we can note that such a variation has yet been obtained from some very general arguments in line with quantum cosmology and with dimensional considerations \cite{Chen90} or by postulating the invariance of equations under a change of scale  \cite{Canuto77}. It has also been \textit{postulated} to explore its consequences as did, for example, Berman  
 \cite{Berman91} who made the hypothesis that $\Lambda(t) = B\, t^{-2}$ and $\rho(t)= A \, t^{-2}$ (leading then when using eq.(\ref{c}) to $\Lambda(t) = B\, a^{-2}$ and $\rho(t)= A \, a^{-2}$ and consequently to some of our results). Fahr and Heyl \cite{Fahr07b} also made the assumption that the total mass density of the universe (matter and vacuum) scales with $a^{-2}$ and find the relation $c=\dot a(t)$ in the particular case $k=0$. They then show that such a scaling abolishes the horizon problem and that the cosmic vacuum energy density can then be reconcilied with its theoretical expected value. 
 Others postulated the Mach's principle or, as did $\ddot{O}$zer  \cite{Ozer87}, made the assumption that the equality  $\rho_M=\rho_c$ is a time-independant feature of the universe from which they deduce $\Lambda \sim a(t)^{-2}$.
 
\section{Conclusion}
We have reconsidered the Hubble diagram of distance moduli and redshifts as obtained from recent observations of type Ia supernovae and of $\gamma$ ray bursts.\\
Traunmüller \cite{HT} and Marosi \cite{Marosi} found independantly that the best equations to account for these observations over the whole range $0.01\,<\,z\,<\,8.1$ can be written (\ref{lum}) and (\ref{muMarosi}) respectively. Our study shows that the standard flat $\Lambda CDM$ model cannot account exactly for these observations neither over the whole range $0.01<z<8.1$ of $z$ values nor over a restricted range as $0.01<z<0.9$, whereas the light model - with eq.(\ref{c}) - is consistent with all the available data 
and matches \textit{perfectly} observations by giving exactly equations (\ref{lum}) and (\ref{muMarosi}).
To conclude, three points can be underlined: 

- eq.(\ref{c}) leads to the best fit curves of observations by using the only parameter $H_0$ that is without needing the knowledge of $\Omega_M$ and $\Omega_{\Lambda}$ and thus without having to adjust any parameter. 

- eq.(\ref{c}) leading to a constant velocity of expansion, all these results shows that models can exist which lead exactly to the observations \textit{without having to consider any acceleration of the expansion of the universe}.

- as briefly shown in the Appendix, introducing eq.(\ref{c}) as an additionnal constraint to solve the Friedmann equations leads to interesting ways to explain several unanswered problems of the standard cosmology concerning, for example, the flatness problem (in the light model, the universe dispays the same evolution as a flat universe and consequenlty must appear to be flat whatever it may be, spherical or not); the horizon problem;  the problem of the observed uniformity in term of temperature and density of the cosmological background radiation
(it is the same tiny part of the early universe that we
observe in any direction around us so that it is quite normal to find the observed background
homogeneity); the small-scale inhomogeneity problem (with the one of the seeds of galaxies and of cosmic structures) and the cosmic coincidence problem  \cite{Vigoureux08},  \cite{Viennot09} \cite{Vigoureux}.
Another appealing feature of this model is also that eq.(\ref{c}) permits to accommodate  simultaneously the equation of state $p_\Lambda  = - \rho_\Lambda c^2$ of the quintessence fluid which generates the cosmological constant (so that it can \textit{perfectly generate} it), with a varying density $\rho_\Lambda \propto a^{-n}$ ($n = 2$ in our case).
It can consequently explain the origin of the cosmological constant with a quintessence fluid which dilutes when the universe expands \cite{Viennot09}.\\
In a more general way, our hypothesis also gives a meaning to some other results: for example, it gives a cosmological meaning to the equation $E = m c^2$: using eq.(\ref{c}), the energy $E = m_0 \,c^2$ of a given rest mass $m_0$ can be seen as a "comoving kinetic energy" ($E = m_0 \,c^2=\alpha \,m_0 \,\dot{R}^2$) of any comoving object carried away by the expansion of the universe.\\
\textit{All theses results have been obtained from the only hypothesis that the speed of light is related to the expansion of the universe (but at the cost of a $a^{-2}$ variation of the scale factor)}. An important feature of eq.(\ref{c}) is thus \textit{its unifying power}. It gives unity to number of results which, for some of them, have yet been obtained by other authors by introducing many quite different, and sometimes \textit{ad hoc}, hypotheses.\\
So, we advocate again the possibility that the universal relations existing between space and time in the so-called "speed of light" and in the expansion of the universe are two aspects of a same phenomenon. 
The constant $c$ was first introduced as the speed of light. However, with the development of
physics, it came to be understood as playing a more fundamental role, its significance being
not directly that of a usual velocity (even though its dimensions are) and one might thus think
of $c$ as being a fundamental constant of the universe. Moreover, the advent of Einsteinian relativity, the fact
that $c$ does appear in phenomena where there is neither light nor any motion (for exemple in
$E = m\,c^2$ or in the relation $c=1/ \sqrt {\epsilon_0\,\mu_0}$) and its double-interpretation in terms of "velocity" of light
and of "velocity" of gravitation forces us to associate it with the theoretical description
of the universe itself rather than that of some of its specific contents.

\section*{Appendix : a brief survey of the model}
\subsection{Friedmann equations and energy-momentum conservation}
Einstein's field equation 
which relates the geometry of space-time to the energy content of the universe can be written (the cosmological constant appears in what follows as a time-dependant function)\\
\begin{equation}\label{einstein}
R_{ij}-\frac{1}{2} R \,g_{ij}= 8 \pi G \,\left(T_{ij} - \frac{\Lambda(t)}{8 \pi G}\,g_{ij}\right)
\end{equation}
c being constant, the derivation of Friedmann's equations is the same as in usual case : taking into account the fact that on very large scale the universe is spatially homogeneous and isotropic to an excellent approximation Einstein's equations reduce to the two Friedmann equations (a dot refers to a derivative with respect to the cosmic time $t$ and we note  $\Lambda(t) = 8 \pi G\, \rho_{\Lambda}(t)$)\\
\begin{equation}\label{Fried1}
\frac{\dot{a}(t)^2}{a(t)^2} =\frac{8 \pi G \rho_M}{3}- \frac{k c^2}{a(t)^2\,R_0^2}+ \frac{\Lambda(t)}{3}=\frac{8 \pi G }{3}(\rho_M+\rho_{\Lambda})- \frac{k c^2}{a(t)^2\,R_0^2}
\end{equation}
and
\begin{equation}\label{Fried2}
\frac{\ddot{a}(t)}{a(t)} =-\frac{4 \pi G}{3}(\rho_M+ 3 \frac{p}{c^2})+ \frac{\Lambda(t)}{3}
\end{equation}
where $G$, $\rho_M$ and $p$ are the gravitationnal constant, matter-energy density and fluid pressure respectively. As usual, the curvature parameter $k$ takes on values $-1, 0, +1$  for negatively curved, flat, and positive curved spatial sections respectively. \\
The only difference between the present model and the standard one comes from the additional constraint (\ref{c}) which expresses a restriction on usual variables characterizing the problem: using $c = \alpha \,R_0\, \dot{a}(t) = const.$ makes $\ddot{a}(t)$ to be $0$ so that Friedmann equations become: 
\begin{equation}\label{Friedc1}
\frac{\dot{a}(t)^2 }{a(t)^2} =\frac{8 \pi G \rho_M}{3 (1+ k \alpha^2)}+ \frac{\Lambda(t)}{3 (1+ k \alpha^2)}
\end{equation}
\begin{equation}\label{Friedc2}
0 =- 4 \pi G\left(\rho_M+ 3 \frac{p}{c^2}\right) +\Lambda(t)
\end{equation}
The energy conservation can be found by differentiation of eq.(\ref{Friedc1}) and by using then eq.(\ref{Friedc2}). It can also be found by writing that the energy-momentum tensor is conserved. Adding to the expression of the energy-momentum tensor for a perfect fluid in the rest frame (diagonal element $(\rho,\,p,\,p,\,p)$) the cosmological term $\rho_{\Lambda} \,g_{\mu \,\nu}$ with diagonal elements $\rho_{\Lambda}(1,\,-1,\,-1,\,-1)$, we get 
\begin{equation}\label{energy}
\dot{\rho}_M + \dot{\rho}_{\Lambda}=-3 \left(\rho_M+\frac{p}{c^2}\right)\frac{\dot{a}(t)}{a(t)}
\end{equation}

\subsection{Solutions of Friedmann equations}
Limiting ourselves to the matter dominated universe ($p(t)= w\, \rho(t)\,=0$) in the case $k=0$ (the general case can be found in \cite{Vigoureux08, Viennot09, Vigoureux}), Friedmann equations (\ref{Friedc1},\,\ref{Friedc2}) lead to :
\begin{equation}\label{rhonow}
\rho_M(t)= \frac{c^2}{4 \pi  G \alpha ^2 R(t)^2}
	\quad \quad	\rho_{\Lambda}(t) 	=\frac{c^2}{8 \pi  G \alpha ^2
   R(t)^2}	\quad \quad
  \rho_c=\frac{3 H(t)^2}{8 \pi  G} =\frac{3 c^2}{8 \pi  G \alpha ^2 R(t)^2}			
\end{equation}
where we indifferently keep the notation $c$ and $R(t)$ for convenience. These results show that the mass density $\rho_M$ of the cosmic fluid and the dark energy density $\rho_{\Lambda}$ scale as $R(t)^{-2}$ and that $\rho_M$, $\rho_{\Lambda}$ and the critical density $\rho_c$ are of the same order of magnitude at all time thus solving the problem of the "cosmic coincidence".
It may be noted that eq.(\ref{rhonow}) leads to $(\Omega_M,\, \Omega_{\Lambda})\simeq (0.66,\,0.33)$ when the standard theory gives $(\Omega_M,\, \Omega_{\Lambda})\simeq (0.3,\,0.7)$ but, again, we must recall that these standard numerical values are not obtained from direct observations and consequently depend in a fundamental way on the theory which is used (see above)\\
It is interesting to note that the $R(t)^{-2}$ variation of $\rho_M$ and $\rho_{\Lambda}$ comes from the presence of the term $\dot{\rho}_{\Lambda}$ in eq.(\ref{energy}). This can be seen by noting that eq.(\ref{rhonow}) implies $\rho_M= 2 \rho_{\Lambda}$ and by introducing this last result into the left-hand side of eq.(\ref{energy}) which becomes
\begin{equation}
\dot{\rho}_{\Lambda}+\dot{\rho}_M=\frac{1}{2}\dot{\rho}_M+\dot{\rho}_M=\frac{3}{2}\dot{\rho}_M
\end{equation}
so that the energy conservation (\ref{energy}) becomes (when $p(t)=0$)
\begin{equation}
\dot{\rho}_M=-2 \rho_M \frac{\dot{a}(t)}{a(t)}
\end{equation}
\textit{where the multiplicating factor $2$ appears instead of $3$} so leading after integrating to $\rho_M(t)\sim a(t)^{-2}$.\\ 
It can also be noted that eq.(\ref{rhonow}) also leads to:
\begin{equation}
M = \frac{4 \pi}{3}a^3 \rho_M =\frac{c^2 (1+\alpha^2)}{3 G \,\alpha^2} a(t)
\end{equation}
showing that the total mass of the universe scales with its cosmic radius.\\
These results call two remarks:\\
- The variation of $\Lambda$ with respect to time and, more precisely, its $a(t)^{-2}$ variation (or, equivalently its $t^{-2}$ variation) in eq.(\ref{rhonow}) has been discussed in \cite{Dolgov83, Ford85, Ratra88} and is suggested by observations  \cite{Overduin98, Chernin00, Baryshev01, Axenides02} for example to solve the coincidence problem.
The $a(t)^{-2}$ variation of $\Lambda$ has also been shown to be in conformity with quantum gravity by Chen and Wu  \cite{Chen90} and to be consistent with the result of \"Ozer  \cite{Ozer87} and other authors  \cite{Khadekar09,Mukhopadhyay11,Ray11} who obtained it in different contexts.\\
- The second remark deals with the scaling of the total mass of the universe with $a(t)$ (this result is preminiscent of black hole the radius of which is also proportional to its mass).
As explained in \cite{Fahr07b}, it has yet been emphasized as possibly true from completely different reasonings by many physicists such as Dirac, Einstein or Hoyle \cite{Einstein17, Dirac37, Whitrow46, Hoyle90, Hoyle92, Fahr07, Fahr07b}.
Moreover it appears, on one hand, that a scaling of masses with the cosmic scale factor is the most natural scale required to make the theory of general relativity conformally scale-invariant (H. Weyl's requirement) and, on the other hand, that it expresses a necessary condition to satisfy  Mach's principle as given by  \cite{Assis94, Brans61}. The possible explanations for such a variation of masses with the scale factor are discussed in \cite{Fahr07}.
\subsection{Some problems of the standard cosmology}
We only summarize here some other results (they can be found with calculations in \cite{Vigoureux08,Viennot09, Vigoureux}):\\
\begin{figure}[!h]
\centering
\includegraphics[scale=0.5]{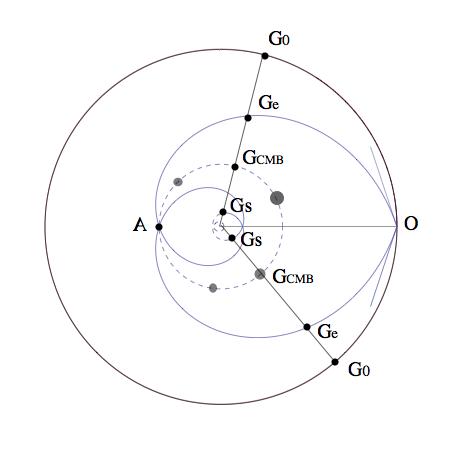}
\caption{The circle of radius $R(t_0)$ represents the universe at the present time  $t_0$. The logarithmic spiral corresponds to the past light cone of the observer $O$, that is, to trajectories of all the light rays that we receive at $t=t_0$. The point $A$ which can be seen in any direction around $O$ can be identified to the "source" of the CMB. The dashed circle corresponds to the universe at time $t_{CMB}$. $A$ represents only a very tiny part of the universe at that time so that, \textit{at that time}, the seeds of galaxies we observe now (points $G_e$) were not at $A$, but here and there on that dashed circle. They are \textit{symbolically} illustrated by grey circles on the dashed circle. Note that they have not the same size. In fact at $t_{CMB}$ the universe did not need to be homogeneous (and was certainly not) so that the seeds of these galaxies at that time could be quite different the ones from the others. The two radius are the world lines of two comoving galaxies: $G_{CMB}$ are galaxies (or their seeds) at time $t_{CMB}$; $G_e$ gives their positions at the time $t_e$ they emitted the light we receive now at $t_0$; $G_0$ are their current (and unknown) positions now. 
\label{fig XX}}
\end{figure}
- the light model shows why the universe is spatially flat at large scale. More precisely, it shows that whatever it may be (flat, spherical or hyperbolical), the universe must appear to be flat when interpetated within the $\Lambda CDM$ model. 
This can be seen by comparing eq.(\ref{Fried1}) with eq.(\ref{Friedc1}): eq.(\ref{Friedc1}) \textit{which describes both flat, closed or open universes following the value of $k$} is quite similar to eq.(\ref{Fried1}) \textit{when we take $k=0$} which then characterizes \textit{a flat universe} in the standard cosmology. A comparison between these two equations thus shows that with eq.(\ref{c}), the universe \textit{must appear to be flat} whatever may be its geometric form (\textit{whatever may be the value of $k$ in eq.(\ref{Friedc1})}) but with more or less matter than expected in the standard model following the value $+1$ or $-1$ of k since the density $\rho$ in eq.(\ref{Fried1}) is changed into $\rho/(1+ k \alpha^2)$ in eq.(\ref{Friedc1}).\\
- the light model is horizon-free. It thus can  explain the observed uniformity in terms of temperature and density of the CMB without having to consider inflation or other hypotheses not only because this allows interactions to homogenize the wole universe but especially because it shows that it is the\textit{ same "element"} of the universe  at $CBR$ time that we see in any directions all around us (see fig.6).\\
- A related comment concerns the problem of the small-scale inhomogeneities needed to produce astronomical structures that are now observed. In the light model, the small fluctuations that we observe now in the CMB \textit{are not those which gave birth to the structures of the universe we can observe}. In fact, as shown on fig.6, the structures which emitted the light we receive at $t=t_0$ were not at $A$ (the event we see all around us) at time $t_A=t_{CMB}$ (and consequently \textit{their seeds were not in the CMB we observe}) but on the circle of radius $ct_A=c t_{CMB}$ which represents the universe at time $t_{CMB}$ (dashed circle in fig.(6)). 
Nothing then imposes that inhomogeneities of the universe at that time (that is on the dashed circle in fig.(6)) be the same as those observed in its very tiny part $A$ (that is to say in the CMB). We cannot know others regions (other than $A$) of the circle of radius $ct_A=c t_{CMB}$ and they may be have overdense parts.\\
-  As said int conclusion, one appealing feature of this model is that eq.(\ref{c}) also permits to accommodate  simultaneously the equation of state $p_\Lambda  = - \rho_\Lambda c^2$ of the quintessence fluid which generates the cosmological constant (so that it can \textit{perfectly generate} it), with a varying density $\rho_\Lambda \propto a^{-n}$ ($n = 2$ in our case).
 It can consequently explain the origin of the cosmological constant with a quintessence fluid which dilutes when the universe expands \cite{Viennot09}.\\

\end{document}